\documentclass[10pt]{article}
\usepackage{multicol}
\usepackage{amssymb}
\usepackage{amsfonts,amssymb,amsmath,amsthm}
\usepackage{epsfig}
\usepackage{slashed}
\usepackage{graphicx}
\usepackage{float}
\usepackage{makecell}
\usepackage{mathrsfs}
\usepackage{bbm}
\usepackage{cite}
\usepackage{times, mathptmx}
\usepackage{enumerate}
\usepackage{longtable}
\usepackage{booktabs}
\usepackage[utf8]{inputenc}
\usepackage{color}
\usepackage[dvipsnames, svgnames, x11names]{xcolor}
\usepackage[colorlinks,linkcolor=NavyBlue,urlcolor=NavyBlue,anchorcolor=NavyBlue,citecolor=NavyBlue]{hyperref}
\usepackage[center]{titlesec}
\renewcommand\thesection{\Roman{section}}
\renewcommand\thesubsection{\Alph{subsection}}

\renewcommand{\thefootnote}{\fnsymbol{footnote}}

\titleformat{\section}
{\centering\normalsize\bfseries}
{\thesection.}
{3pt}
{}
\titleformat{\subsection}
{\centering\normalsize\bfseries}
{\thesubsection.}
{3pt}
{}

\begin{document}

\begin{center}
{\large \bf Null geodesics, quasinormal modes and the correspondence with shadows in high-dimensional Einstein-Yang-Mills spacetimes}
\vspace{4mm}

{{ Yang Guo\footnote{\em E-mail: guoy@mail.nankai.edu.cn}
and Yan-Gang Miao}\footnote{\em Corresponding author E-mail: miaoyg@nankai.edu.cn}

\vspace{1mm}
{\normalsize \em School of Physics, Nankai University, Tianjin 300071, China}}
\end{center}

\noindent
Null geodesics, quasinormal modes of a massless scalar field perturbation and the correspondence with shadow radii are investigated in the background spacetime of high-dimensional Einstein-Yang-Mills black holes. Based on the properties of null geodesics, we obtain the connection between the radius of a photon sphere and the radius of a horizon in the five- and six-dimensional Einstein-Yang-Mills spacetimes. Especially in the five-dimensional case, there exist two branches for the radius of a photon sphere, but only the branch outside the event horizon satisfies the condition of circular null geodesics. Moreover, we find no reflecting points of shadow radii and no spiral-like shapes on the complex plane of quasinormal frequencies 
and verify the correspondence between the quasinormal modes in the eikonal limit and shadow radii  in high-dimensional Einstein-Yang-Mills spacetimes.

\vspace{5mm}

\renewcommand{\thefootnote}{\arabic{footnote}}
\setcounter{footnote}{0}
\setcounter{page}{2}
\pagenumbering{arabic}

\section{Introduction}
Quasinormal modes (QNMs) are usually used to depict the stability of black holes that are perturbed by an external field or by the metric of black hole spacetimes, and they also contain the information of gravitational waves. The successful detection of gravitational waves by the LIGO Scientific and Virgo Collaborations~\cite{GW} opens a new window for the future astronomy, the so-called gravitational wave astronomy.
In addition, the images of M87 observed by the Event Horizon Telescope Collaboration~\cite{EHT} show the shadow of the supermassive black hole. The two remarkable achievements mark the beginning of the new era of gravitational wave astronomy, which in turn greatly facilitates the studies of QNMs and shadow radii and their correspondence.

Based on the general relativity and alternative theories of gravity, QNMs have been studied in a wide range of issues~\cite{SHM,RAK,RAKZ,SYG,IS,CLZWJJ}. Most of the research of QNMs rely on numerical computations, but only a few on analytical computations. For the latter, the lowest QNM frequencies of a vector-type field perturbation have been computed~\cite{AS} analytically for a generic black hole with a translationally invariant horizon, where the QNM frequencies approach~\cite{LM} a large imaginary number plus $\ln 3/(8\pi G_{N}M)$. Moreover, an analytic computation of highly damped QNMs has been made~\cite{LMN} in term of a monodromy technique. 

Now we make a brief review on the correspondence between QNMs and other important issues, such as spacelike geodesics, phase transitions, and null geodesics, etc., which naturally gives our motivation and aim of the present paper.
It has been founded~\cite{LFHKS,GFL} that the QNM frequencies in the large black hole mass limit are determined by the spacelike geodesics with the boundary of the Penrose diagram, based on which the quantum aspect of gravity behind horizons can be probed in the context of the gauge/gravity duality. For the four-dimensional Reissner-Nordstr\"om (RN) black hole, a spiral-like shape has been shown~\cite{JJP} on the complex plane of QNM frequencies when the black hole evolves to its second order phase transition point. This relation between QNMs and phase transitions gives an opportunity to probe~\cite{XHCWCL} the thermodynamics and dynamics of black holes. Moreover, it has been proposed~\cite{VCMBWZ} that the real and imaginary parts of QNMs in the eikonal limit have a compact connection to the angular velocity and Lyapunov exponent of unstable circular null geodesics. Such a correspondence was later extended~\cite{IZSYG} to investigate the connection between QNMs and the gravitational lensing. However, it has been pointed out~\cite{RAKS} for the Einstein-Lovelock theory that the QNM frequencies determined by the angular velocity and Lyapunov exponent deviate from those by the WKB method, which implies that the correspondence between QNMs and null geodesics is violated in the Einstein-Lovelock theory. 

Recently, the correspondence between the real part of QNMs in the eikonal limit and shadow radii of black holes has been suggested~\cite{KJ1912,KJ2004} and applied to 
a black hole spacetime surrounded by the perfect fluid dark matter, where an interesting discovery is the existence of a reflecting point of shadow radii that corresponds to the maximal value of the real part of QNM frequencies. The connection between QNMs and shadow radii may, alternatively, provide a physical picture at the semi-classical level that the gravitational waves can be understood as such a phenomenon that a massless particle propagates along an outmost and unstable orbit of null geodesics and spreads transversely out to infinity. 
Based on the correspondence of Refs.~\cite{KJ1912,KJ2004} plus a sub-leading contribution derived from the WKB method~\cite{SIW},    
a modified correspondence between QNMs and shadow radii has been proposed~\cite{BCMFO} and verified to be a good agreement with the WKB method in the eikonal limit for a $D$-dimensional Tangherlini black hole and a four-dimensional spherically symmetric black hole surrounded by anisotropic fluids. 
As stated in our previous work~\cite{YGM}, the Einstein-Yang-Mills (EYM) theory~\cite{SHMH} is more challenging in finding analytic solutions than the Einstein-Maxwell theory due to the former's intrinsic properties related to QNMs and phase transitions. Therefore, 
we are interested in whether the modified correspondence between QNMs in the eikonal limit and shadow radii works well or not and also eager for probing the related issues, such as whether a reflecting point of shadow radii and a spiral-like shape on the complex  plane of QNM frequencies exist or not in the context of the EYM theory, which aims at our main purpose in this paper. That is, we shall investigate such a  correspondence between QNMs in the eikonal limit and shadow radii by dealing with the circular null geodesics of a massless particle around the EYM black holes with the $SO(D-1)$ gauge group. The significance of our issue lies in richening the relationship between gravitational waves and shadows of black holes, the two amazing achievements in the new century.   

The outline of the present paper is as follows. In Sec.~\ref{sec:GEO} we investigate the circular null geodesics of the EYM spacetimes in $D=5$ and $D=6$ as an example of $D>5$ to find the connection between photon sphere radii and horizon radii. In order for this paper to be self-contained, we review shortly in Sec.~\ref{sec:QNM} the derivation of the perturbation equations of a massless scalar field and the calculation of the corresponding QNMs for different values of multiple numbers and overtone numbers by the improved WKB method in a static spherically symmetric EYM spacetime. 
In Sec.~\ref{sec:Cor} we focus on the correspondence between the QNMs in the eikonal limit and shadow radii and the related issues.
Finally, we give our conclusions in Sec.~\ref{sec:conclusions}. We adopt the geometric unit throughout this paper as usual.

\section{Null geodesics}\label{sec:GEO}

For a $D$-dimensional static and spherically symmetric spacetime, the metric can be described by

\begin{eqnarray}
{\rm d}s^2=-f(r){\rm d}t^2+\frac{{\rm d}r^2}{f(r)}+r^2{\rm d}\Omega_{D-2}^2,\label{metric}
\end{eqnarray}
where ${\rm d}\Omega_{D-2}^2$ is the line element of the unit sphere $S^{D-2}$ with the usual angular coordinates $\theta_{i}\in[0,\pi]$, $i=1, 2, \dots, D-3$, and $\varphi\in[0,2\pi]$. A freely falling massless particle moving along a null geodesic satisfies the equation,
\begin{eqnarray}
g_{\mu\nu}\frac{{\rm d}x^{\mu}}{{\rm d}\lambda}\frac{{\rm d}x^{\nu}}{{\rm d}\lambda}=0,\label{GEO}
\end{eqnarray}
where the greek indices, $\mu, \nu=0, 1, \dots, D-1$, describe the $D$-dimensional spacetime, and $\lambda$ is an affine parameter of the null geodesic.
In order to give the condition satisfied by circular null geodesics in the above static and spherically symmetric spacetime,  Eq.~(\ref{metric}), we shall consider a free massless particle orbiting in the equatorial hyperplane ($\theta_{i}=\pi/2$). Without loss of generality, one has the Lagrangian of a massless particle,
\begin{eqnarray}
\mathcal{L}=\frac{1}{2}\left[ -f(r)\dot{t}^2+\frac{\dot{r}^2}{f(r)}+r^2\dot{\varphi}^{2}\right], 
\end{eqnarray}
where the dot stands for the differentiation with respect to an affine parameter, and there exist two Killing vector fields $\partial/\partial t$ and $\partial/\partial\varphi$ in this spacetime. Correspondingly, one obtains the energy and the angular momentum of the massless particle,
\begin{eqnarray}
E=f(r)\dot{t},\qquad L=r^2\dot{\varphi}.\label{ergmom}
\end{eqnarray}
Given these two conserved quantities, one can rewrite the null geodesic equation as follows,
\begin{eqnarray}
\dot{r}^2=V(r),
\end{eqnarray}
with the effective potential
\begin{eqnarray}
V(r)=E^2-\frac{L^2}{r^2}f(r).\label{poten1}
\end{eqnarray}
For a circular null geodesic, the effective potential satisfies~\cite{JMBPT} the following conditions,
\begin{eqnarray}
V(r)=0,\qquad \frac{dV(r)}{dr}=0,\qquad \frac{d^2V(r)}{dr^2}>0,\label{cond}
\end{eqnarray}
which can be used to determine the radius of a photon sphere and the instability of the bound circular orbits. 

For the five-dimensional Einstein-Yang-Mills black hole~\cite{SHMH}, its metric takes the form,
\begin{eqnarray}
	f(r)=1-\frac{M}{r^2}-\frac{2Q^2}{r^2}\ln(r), \label{5EYM-metric}
\end{eqnarray}
where $M$ denotes the black hole mass and $Q$ the only non-zero gauge charge, and its horizon has two branches as follows:
\begin{eqnarray}
r_{-}=\exp\left\{-\frac{1}{2Q^2}\left[M+Q^2W_{0}\left(-\frac{1}{Q^2}\exp\left(-\frac{M}{Q^2}\right)\right)\right] \right\},\label{ihori5}
\end{eqnarray}
\begin{eqnarray}
r_{+}=\exp\left\{-\frac{1}{2Q^2}\left[M+Q^2W_{-1}\left(-\frac{1}{Q^2}\exp\left(-\frac{M}{Q^2}\right)\right)\right] \right\},\label{ehori5}
\end{eqnarray}
where $r_{-}$ means the Cauchy horizon radius and $r_{+}$ the event horizon radius. Here $W_k(x)$, $k=0, \pm 1, \pm 2, \dots$, are called Lambert's W functions~\cite{RMCGHJK}, where $W_0(x)$ is referred to as the principal branch.
By using Eq.~(\ref{cond}), 
we derive the two branches of the radius of a photon sphere,\footnote{At first, we obtain the two solutions from $\left.\frac{dV(r)}{dr}\right|_{r=r_{\rm ps}}=0$, and then verify whether they satisfy $V(r_{\rm ps})=0$ and  $\left.\frac{d^2V(r)}{dr^2}\right|_{r=r_{\rm ps}}>0$.} 
\begin{eqnarray}
r_{\rm {ps}}^{-}=\exp\left\{-\frac{1}{4Q^2}\left[2M+2Q^2 W_0\left(-\frac{1}{2Q^2}\exp\left(\frac{Q^2-2M}{2Q^2}\right)\right)-Q^2\right] \right\},
\end{eqnarray}
\begin{eqnarray}
r_{\rm {ps}}^{+}=\exp\left\{-\frac{1}{4Q^2}\left[2M+2Q^2W_{-1}\left(-\frac{1}{2Q^2}\exp\left(\frac{Q^2-2M}{2Q^2}\right)\right)-Q^2\right] \right\}.\label{psr5}
\end{eqnarray}

For the six-dimensional EYM black hole~\cite{SHMH}, its metric has the form,
\begin{eqnarray}
f(r)=1-\frac{M}{r^3}-\frac{3Q^2}{r^2}, \label{6EYM-metric}
\end{eqnarray}
and its horizon has only one branch, i.e., the event horizon,
\begin{eqnarray}
r_{+}=\frac{1}{2}\left(4M+4\sqrt{M^2-4Q^6}\right)^{1/3}+\frac{2Q^2}{\left(4M+4\sqrt{M^2-4Q^6}\right)^{1/3}}.\label{hori6}
\end{eqnarray}
Similarly, we derive only one radius of a photon sphere by using Eq.~(\ref{cond}),
\begin{eqnarray}
r_{\rm {ps}}^{+}=\frac{1}{2}\left(10M+2\sqrt{25M^2-128Q^6}\right)^{1/3}+\frac{4Q^2}{\left(10M+2\sqrt{25M^2-128Q^6}\right)^{1/3}}.\label{psr6}
\end{eqnarray}

\begin{figure}[H]
	\centering
	\includegraphics[width=0.6\linewidth]{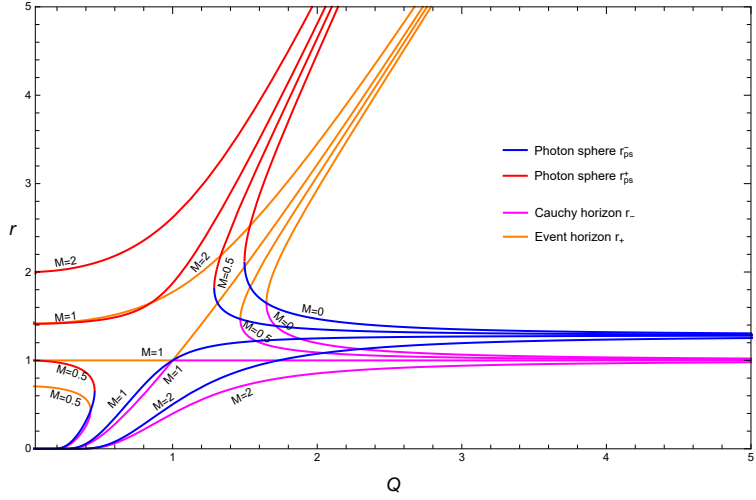}  
	\caption{The radii of photon spheres and radii of horizons on the $Q-r$ plane for four different values of black hole masses, $M=0, 0.5, 1, 2$, in the five-dimensional EYM spacetime.
	} \label{r-5d}
\end{figure}

\begin{figure}[H]
	\centering
	\includegraphics[width=0.6\linewidth]{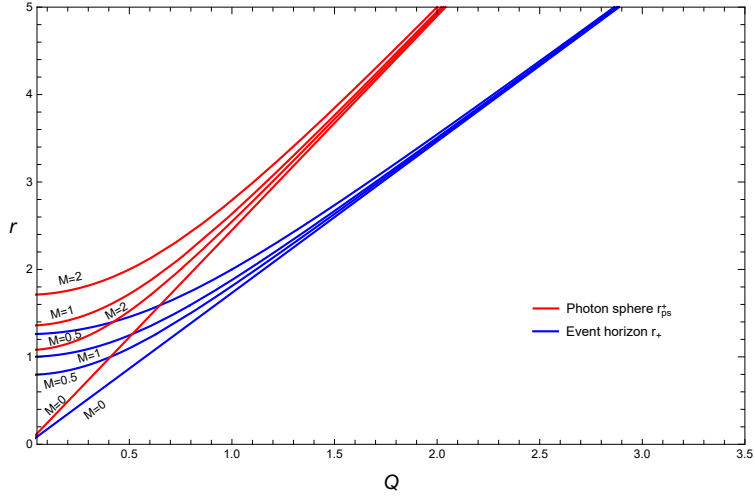}  
	\caption{The radii of photon spheres and radii of horizons on the $Q-r$ plane for four different values of black hole masses, $M=0, 0.5, 1, 2$, in the six-dimensional EYM spacetime.
	} \label{r-6d}
\end{figure}

For the sake of intuition, we plot the radii of photon spheres and horizons with respect to the gauge charge $Q$ in Fig.~\ref{r-5d} and Fig.~\ref{r-6d} in which we can see the connection between the radii of photon spheres and the radii of horizons.

For the five-dimensional EYM black hole, we can see from Fig.~\ref{r-5d} that the position of one branch (blue line for $r_{\rm {ps}}^{-}$) of photon sphere radii is located between the Cauchy horizon (purple line for $r_-$) and the event horizon (orange line for $r_+$), while the position of the other (red line for $r_{\rm {ps}}^{+}$) is outside the event horizon. In addition, we find that the branch of photon sphere radii ($r_{\rm {ps}}^{-}$) does not satisfy the first and third conditions of Eq.~(\ref{cond}). Such  properties of photon sphere radii are consistent with the assumption~\cite{YDFR} that there exists one photon sphere whose radius satisfies the range of values, $r_{\rm {ps}}^{+} \in (r_{+},\infty)$, which supports the existence of unstable circular null geodesics in the five-dimensional EYM spacetime. 

For the six-dimensional EYM black hole, we can see from Fig.~\ref{r-6d} that there is only one photon sphere whose radius (red line for $r_{\rm {ps}}^{+}$) is located outside the event horizon (blue line for $r_+$). 

As a common property of the five- and six-dimensional EYM spacetimes, it is clear from the two figures that both the radius of a photon sphere and the radius of a horizon are mass independent for a large value of the charge.

\section{Quasinormal modes of a massless scalar field perturbation}\label{sec:QNM}

The massless scalar field $\Phi$ propagating in a curved spacetime is described by the following equation,
\begin{eqnarray}
	\frac{1}{\sqrt{-g}}\partial_{\mu}\left(\sqrt{-g}g^{\mu\nu}\partial_{\nu}\Phi\right)=0, \label{KGEq.}
\end{eqnarray}
where $g^{\mu\nu}$ denotes the inverse of $g_{\mu\nu}$ and $g$ the determinant of $g_{\mu\nu}$, respectively. Substituting the following decomposition of variables,
\begin{eqnarray}
\Phi(t,r,\Theta,\varphi)=\sum_{l,m}e^{-i\omega t}\frac{\psi(r)}{r^{(D-2)/2}}Y_{lm}(\Theta,\varphi),
\end{eqnarray}
into Eq.~(\ref{KGEq.}), where $\Theta$ stands for $\theta_1, \theta_2, \dots, \theta_{D-3}$ and $Y_{lm}(\Theta, \varphi)$ the spherical harmonics of $D-2$ degrees, and defining the ``tortoise" coordinate by the relation, ${\rm d}r_*={\rm d}r/f(r)$, we get the following radial equation in its standard form, 
\begin{eqnarray}
\left[ \partial_{r_*}^2+\omega^2-{\cal V}(r)\right] \psi(r)=0.\label{ODE}
\end{eqnarray}
The QNMs as the solution of the differential equation satisfy the following boundary conditions,
\begin{eqnarray}
\psi\sim e^{-i\omega(t\mp r_{*})},\qquad r_*\to\pm\infty,
\end{eqnarray}
and oscillate and decay at a complex frequency, 
\begin{eqnarray} 
\omega=\omega_{\rm R}-i\omega_{\rm I}.
\end{eqnarray}
In the EYM spacetimes, the effective potentials of the perturbation field\footnote{The two effective potentials are defined at the semi-classical level, while the effective potential of Eq.~(\ref{poten1}) has its meaning at the classical level, and the former turns back~\cite{WSWL} to the latter in the eikonal limit.} take~\cite{YGM} the forms,
\begin{eqnarray}
{\cal V}(r)=\left(1-\frac{M}{r^2}-\frac{2Q^2\ln(r)}{r^2}\right)\left[ \frac{4l(l+2)+3}{4r^2}+ \frac{9M-12Q^2}{4r^4}+\frac{3Q^2\ln(r)}{2r^4} \right],   \qquad{D=5},\label{EYM-potential}
\end{eqnarray}
\begin{eqnarray}
{\cal V}(r)=\left(1-\frac{M}{r^{D-3}}-\frac{(D-3)Q^2}{(D-5)r^2}\right)\bigg[\frac{4l(l+D-3)+(D-2)(D-4)}{4r^2}+\frac{(D-2)^2M}{4r^{D-1}}  \nonumber \\
-\frac{(D-2)(D-3)(D-8)Q^2}{4(D-5)r^4}   \bigg], \qquad{D>5}.\label{EYM-potential2}
\end{eqnarray}

\begin{figure}[H]
	\centering
	\includegraphics[width=0.45\linewidth]{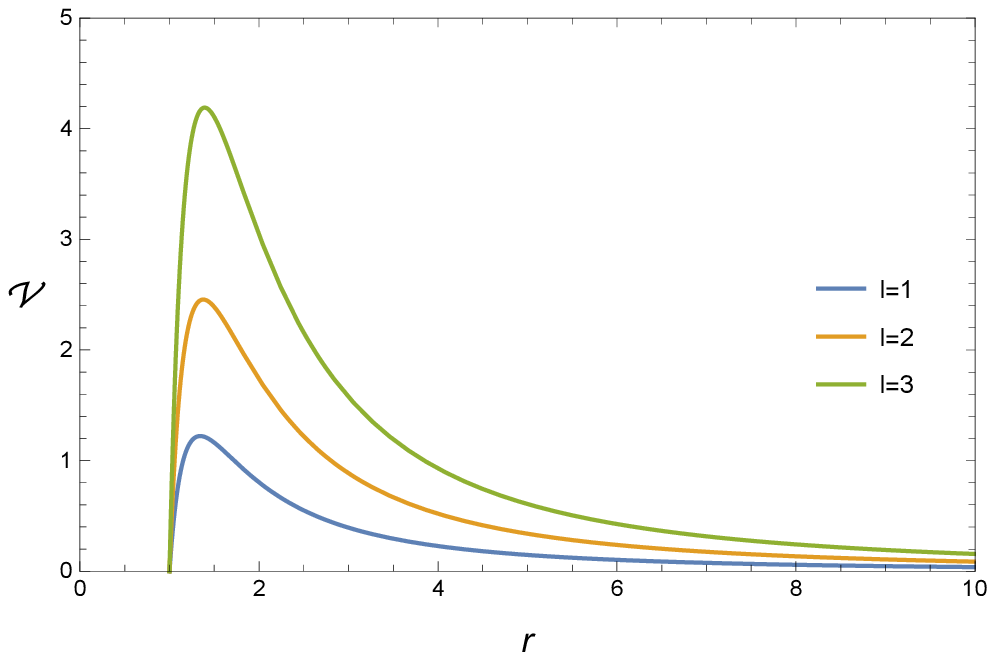} \qquad 
	\includegraphics[width=0.45\linewidth]{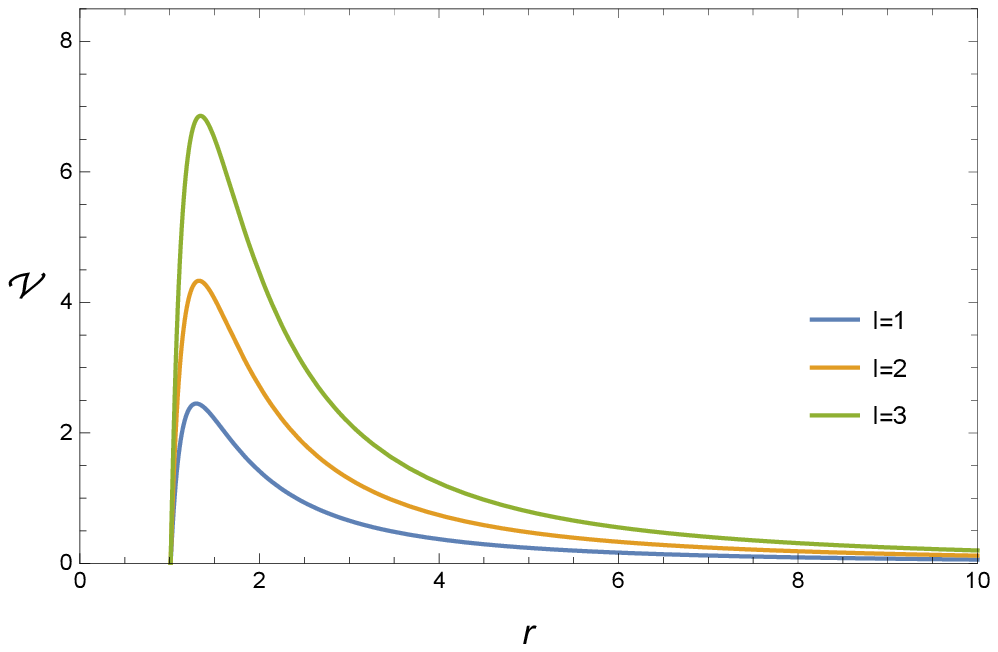} 
	\caption{The effective potential with respect to the radial coordinate in the five-dimensional EYM black hole (left) and in the six-dimensional EYM black hole (right) for the fixed charge and mass, $Q=0.1$ and $M=1$, but different values of multiple numbers.
	} \label{fig:Vr}
\end{figure}

Now we analyze the behavior of the effective potentials for a changing multiple number $l$ (also called the angular quantum number). 
We fix the gauge charge $Q$ and the black hole mass $M$, and then plot Fig.~\ref{fig:Vr} for the effective potentials with respect to the radial coordinate. We find that the effective potentials become large for a big multiple number   
in both the five- and six-dimensional cases. Moreover, we have calculated the expected QNM frequencies in our previous work~\cite{YGM} by using  the improved WKB approximation\cite{JMO,RAKZZ}.
As we shall see in the next section, the behaviors of the effective potentials and QNM frequencies for a changing multiple number $l$  will influence the correspondence between QNMs and shadow radii in high-dimensional EYM black holes. 

\section{Correspondence between QNMs and shadow radii}\label{sec:Cor}
In general, the shadow shape of a black hole depends on whether the black hole rotates or not. For a static and spherically symmetric black hole,\footnote{For the evolution of photon spheres in dynamical black hole spacetimes and the connection to shadows, please refer to Ref.~\cite{MCS}.} as the EYM black holes we are dealing with, the shadow has the spherical symmetry and is described by a photon sphere. From the definition of a shadow radius, or directly deriving  from Eq.~(\ref{cond}), one can write~\cite{VPTBK} the shadow radius in term of the radius of a photon sphere,
\begin{eqnarray}
	\left. R_{\rm {sh}}=\frac{r}{\sqrt{f(r)}} \right|_{r=r_{\rm{ps}}^+}.\label{shrad}
\end{eqnarray}
As a result, we compute the shadow radii in the five- and six-dimensional EYM black holes, respectively, by substituting Eqs.~(\ref{5EYM-metric}), (\ref{psr5}), (\ref{6EYM-metric}) and (\ref{psr6}) into Eq.~(\ref{shrad}), 
\begin{eqnarray}
R_{\rm {sh}}=\frac{\exp\left[-\frac{1}{2Q^2}\left(M+Q^2\Xi\right)-Q^2 \right]}{\sqrt{1-(M+2Q^2)\left[-\frac{1}{2Q^2}\left(M+Q^2\Xi\right)-Q^2 \right]\exp\left[\frac{1}{Q^2}\left(M+Q^2\Xi\right)+2Q^2 \right]}},\label{R5} \qquad D=5,
\end{eqnarray}
\begin{eqnarray}
R_{\rm {sh}}=\frac{\frac{1}{2}\Re^{1/3}+\frac{4Q^2}{\Re^{1/3}}}{\sqrt{1-\frac{M}{\left( \frac{1}{2}\Re^{1/3}+\frac{4Q^2}{\Re^{1/3}} \right) ^3}-\frac{3Q^2}{\left( \frac{1}{2}\Re^{1/3}+\frac{4Q^2}{\Re^{1/3}}\right) ^2}}},\label{R6} \qquad D=6,
\end{eqnarray}
where
\begin{eqnarray}
\Xi\equiv W_{-1}\left(-\frac{1}{2Q^2}\exp\left(\frac{Q^2-2M}{2Q^2}\right)\right),\label{defxi}
\end{eqnarray}
\begin{eqnarray}
\Re\equiv 10M+2\sqrt{25M^2-128Q^6}.\label{delcalr}
\end{eqnarray}

\begin{figure}[H]
	\centering
	\includegraphics[width=0.45\linewidth]{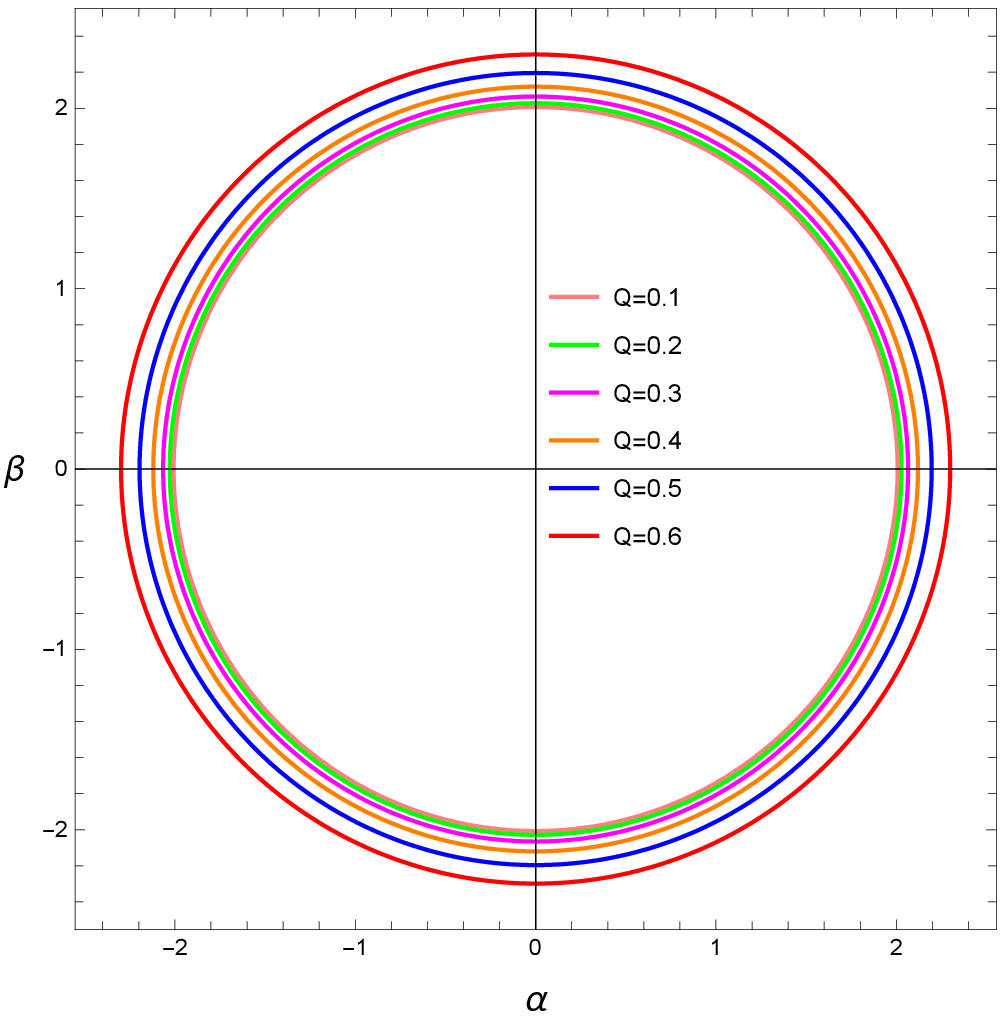} \qquad 
	\includegraphics[width=0.45\linewidth]{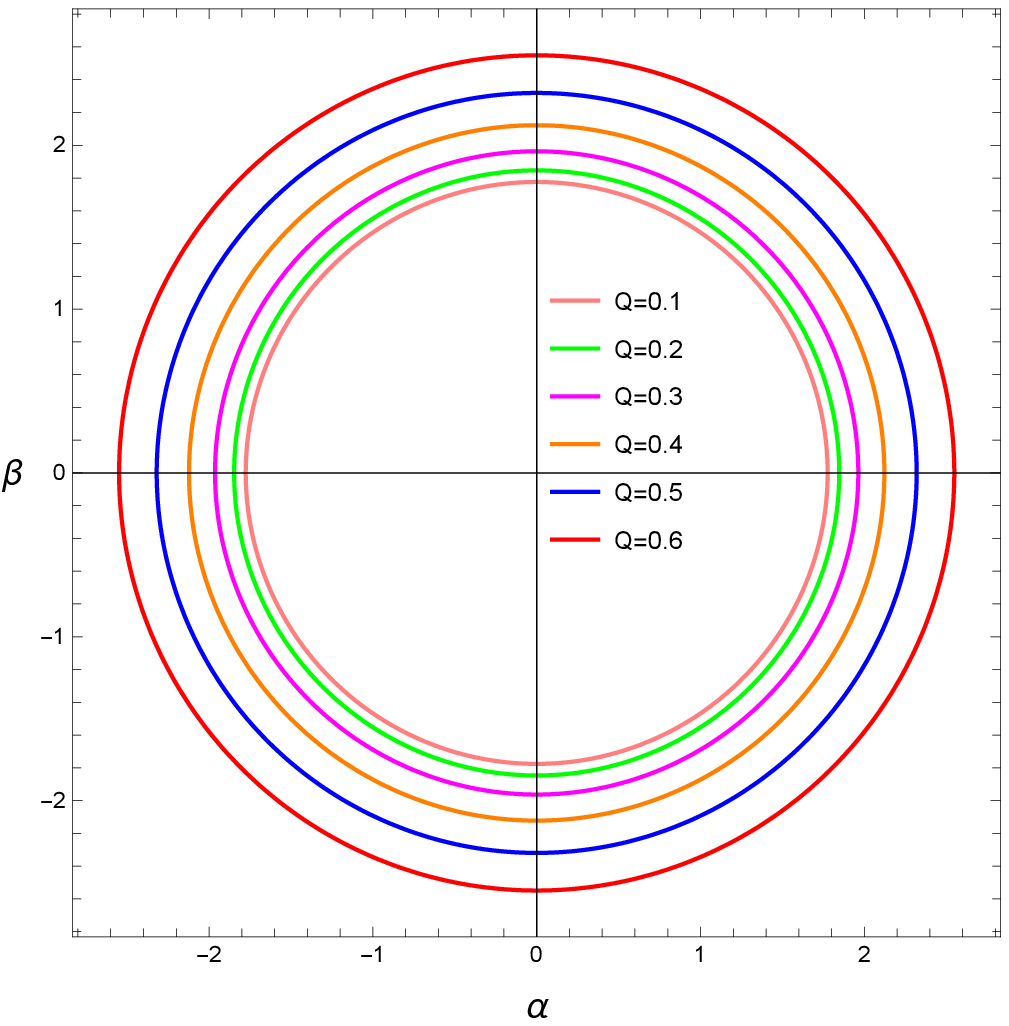} 
	\caption{The profile of shadows cast by the five-dimensional EYM black hole (left) and the six-dimensional EYM black hole (right) for six different values of the gauge charge, $Q=0.1, 0.2, 0.3, 0.4, 0.5, 0.6$. 
	} \label{fig:shadow}
\end{figure}

For a spherically symmetric spacetime, the shadows can be plotted~\cite{SEVE} intuitively via the celestial coordinates $(\alpha,\beta)$  defined~\cite{KJ2004} by
\begin{eqnarray}
\alpha&\equiv&-\xi \csc\theta_0, \\
\beta&\equiv&\pm\sqrt{\eta+a^2\cos^2\theta_0-\xi^2\cot^2\theta_0}\,, 
\end{eqnarray}
where $\theta_0$ denotes the inclination angle of the observer, $a$ the black hole spin, $\xi\equiv L/E$, $\eta\equiv K/E^2$, and $K$ is the Carter constant. It is easy to verify that the celestial coordinates have a close connection with the shadow radius,
\begin{eqnarray}
\sqrt{\alpha^2+\beta^2}=R_{\rm {sh}}.
\end{eqnarray}
Thus the shadows cast by the five- and six-dimensional EYM black holes can be shown in terms of the celestial coordinates in Fig.~\ref{fig:shadow}, where the mass parameter is set to be unit, $M=1$. 

It is clear from Fig.~\ref{fig:shadow} that the shadow radii increase in the two black holes when the gauge charge increases and that the increments of the shadow radii also increase for an increasing charge but with the same interval, e.g., $\Delta Q=0.1$. Moreover, 
the shadow span in the six-dimensional EYM black hole is larger than that in the five-dimensional EYM black hole when the gauge charge is increasing from $0.1$ to $0.6$. 
The other property that is worth noting is that there are no reflecting points\footnote{The reflecting point exists~\cite{KJ1912} in the spacetime of black holes surrounded by the perfect fluid dark matter, where the metric function contains a logarithmic term. Here we emphasize that no such a reflecting point exists in the five-dimensional EYM black hole although a logarithmic term appears in its metric function, see Eq.~(\ref{5EYM-metric}). This is not strange because the two models have obvious differences from each other. For the former, the reflecting point appears when the dark matter parameter increases; while for the latter, it does not appear when the gauge charge increases.} at which the shadow radius will shrink with the increasing of the gauge charge.

\begin{figure}[htbp]
	\centering
	\includegraphics[width=0.45\linewidth]{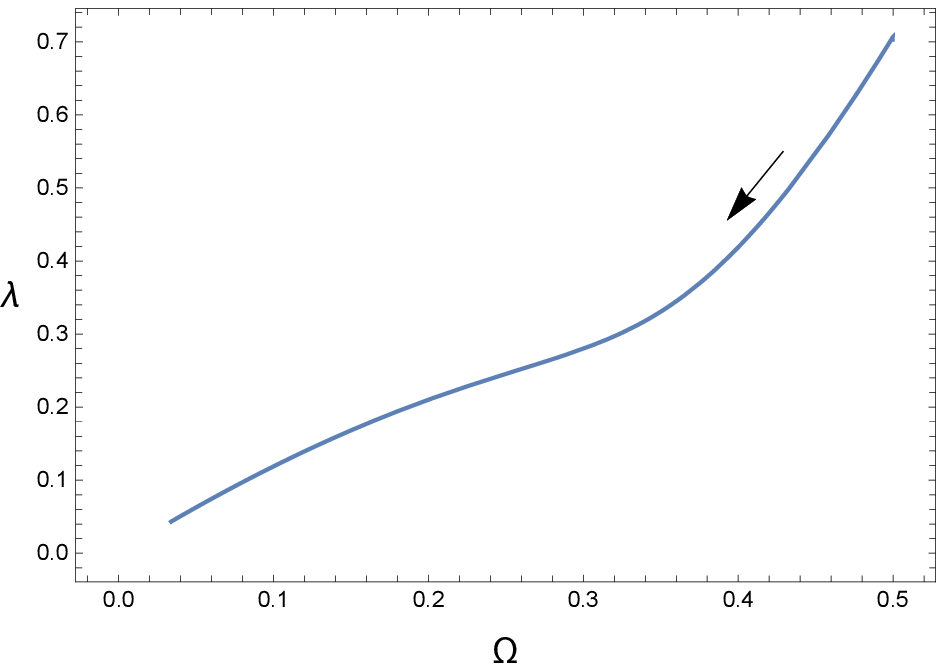} \qquad 
	\includegraphics[width=0.455\linewidth]{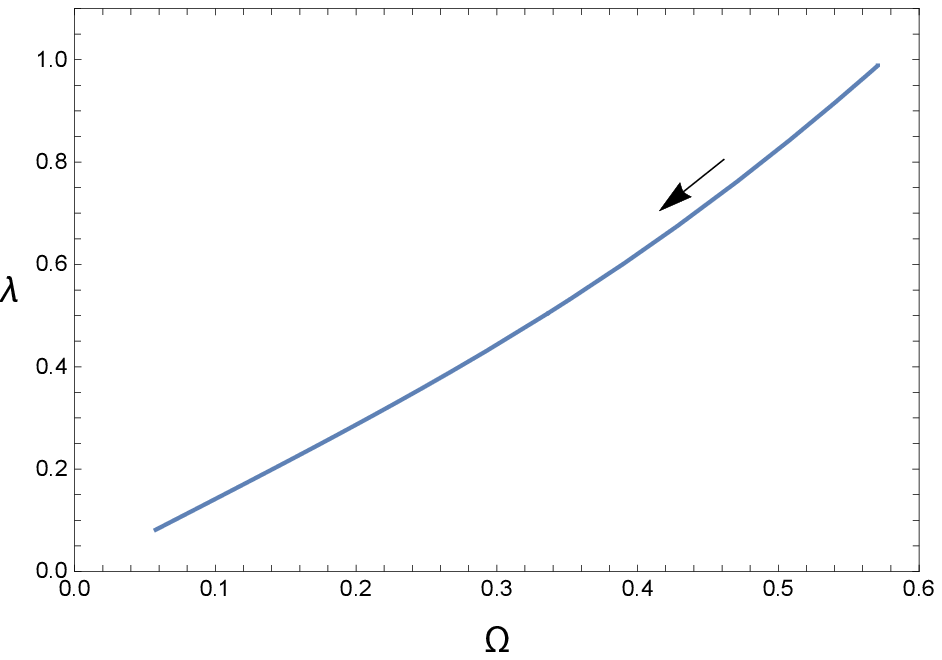} 
	\caption{The angular velocity versus the Lyapunov exponent in the five-dimensional EYM black hole (left) and the six-dimensional EYM black hole (right), where the black arrow denotes the direction of the increasing charge.
	} \label{fig:ol}
\end{figure}

As revealed by the correspondence between QNMs and null geodesics, 
the real and imaginary parts of QNM frequencies in the eikonal limit can be determined~\cite{VCMBWZ} by the angular velocity $\Omega$ and Lyapunov exponent $\lambda$, 
\begin{eqnarray}
	\omega=\Omega l-i\lambda\left(n+\frac{1}{2}\right),\label{QNM}
\end{eqnarray}
where $n$ is called the overtone number, and the angular velocity and Lyapunov exponent can be derived from Eqs.~(\ref{ergmom})-(\ref{cond}), or directly be written~\cite{KJ1912} as the following forms at a photon sphere radius, 
\begin{eqnarray}
\left. \Omega=\frac{\dot{\varphi}}{\dot{t}}=\frac{\sqrt{f(r)}}{r} \right|_{r=r_{\rm{ps}}^+},\label{Omega}
\end{eqnarray}
\begin{eqnarray}
\lambda= \sqrt{\frac{V^{\prime\prime}(r)}{2\dot{t}^2}}=\left.\sqrt{\frac{f(r)[2f(r)-r^2f''(r)]}{2r^2}} \right| _{r=r_{\rm{ps}}^+}.\label{omelam}
\end{eqnarray}
As a result, we can calculate the angular velocity and Lyapunov exponent by substituting Eqs.~(\ref{5EYM-metric}), (\ref{psr5}), (\ref{6EYM-metric}) and (\ref{psr6}) into Eq.~(\ref{Omega}) and Eq.~(\ref{omelam}) for the five- and six-dimensional EYM black holes,  and then plot the $\Omega - \lambda$ graph in Fig.~\ref{fig:ol} by choosing  a very wide range of values of the gauge charge from $0.0$ to $5.0$ in order not to miss a reflecting point and a spiral-like shape, where the mass is set to be unit, $M=1$.

It is quite obvious that the angular velocity and Lyapunov exponent of the null geodesics are decreasing monotonically with the increasing of the gauge charge from $0.0$ to $5.0$. 
Considering the correspondence between QNMs and shadow radii given in Refs.~\cite{KJ1912,KJ2004}, the shadow radius is inversely proportional to the angular velocity in the eikonal limit, $R_{\rm sh}\propto \frac{1}{\Omega}$,
we can deduce that the shadow radius does not shrink with the increasing of the gauge charge, that is, no reflecting points appear, which is clear from the monotonic behavior of the angular velocity in Fig.~\ref{fig:ol}. In addition, the monotonic behavior of the Lyapunov exponent indicates that there are no spiral-like shapes\footnote{A recent work has shown~\cite{WSWL} that the spiral-like shape exists only in the four-dimensional RN black hole, but does not in high-dimensional RN black holes.} in high-dimensional EYM black holes, which implies that the high-dimensional EYM black holes have different phase transition behaviors from those of the four-dimensional RN black hole because the spiral-like shape on the $\Omega-\lambda$ plane influences~\cite{JJP} thermodynamic phase transitions. 

Given the radii of photon spheres, we can determine the QNM frequencies via the angular velocity and Lyapunov exponent by using Eqs.~(\ref{QNM})-(\ref{omelam}) together with Eqs.~(\ref{5EYM-metric}), (\ref{psr5}), (\ref{6EYM-metric}) and (\ref{psr6}). 
However, we find for the EYM black holes that the real components of frequencies determined by the angular velocity are less than those by the improved WKB approximation for a small multiple number.
The reason relies on the fact that the relation between QNMs and null geodesics, Eqs.~(\ref{QNM})-(\ref{omelam}), is valid only at a big multiple number, i.e., in the eikonal limit. 

In order to reduce the deviation of real parts of QNMs, we appeal to a modified correspondence~\cite{BCMFO} between QNMs in the eikonal limit and shadow radii of black holes, 
\begin{eqnarray}
	\omega_{\rm R}={R_{\rm{sh}}^{-1}}\left(l+\frac{D-3}{2}\right),\label{revreal}
\end{eqnarray}
where the terms of equal and higher orders of $l^{-1}$ have been omitted due to the limit $l \gg 1$. In fact, this modified relation was derived~\cite{BCMFO} with the help of the WKB method by the addition of the sub-leading term ${R_{\rm{sh}}^{-1}}{(D-3)}/{2}$ to the leading one ${R_{\rm{sh}}^{-1}}{l}$, and 
verified to coincide with the WKB method at a big multiple number, say, for instance, from $10$ to $10^5$, for a $D$-dimensional Tangherlini black hole and a four-dimensional spherically symmetric black hole surrounded by anisotropic fluids. 
We want to see if such a relation has a good agreement to the improved WKB method at a small multiple number for the EYM black holes.

\begin{figure}[htbp]
	\centering
	\includegraphics[width=0.45\linewidth]{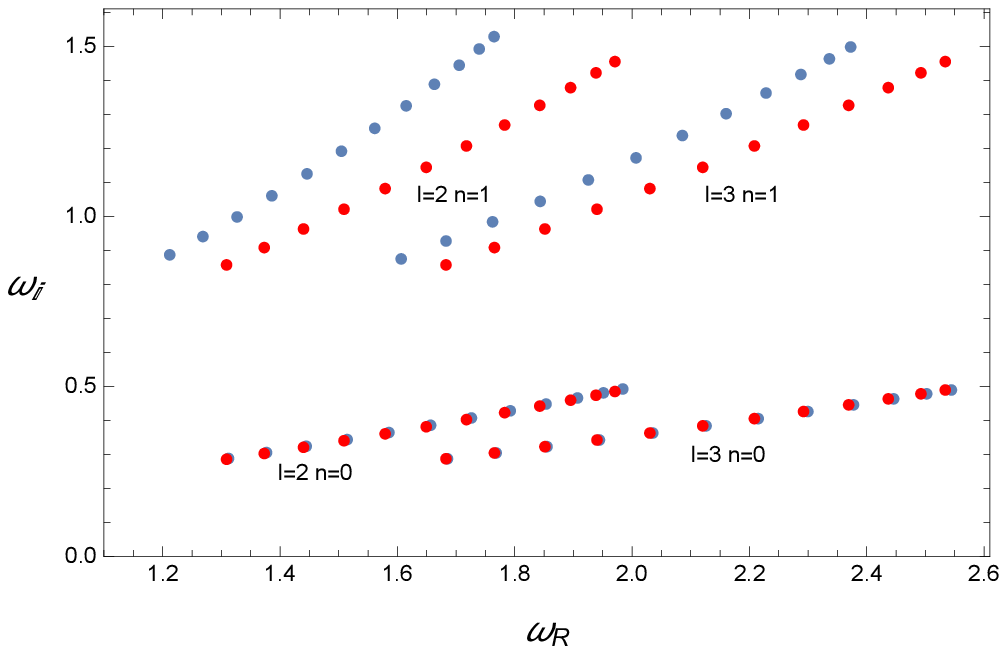} \qquad 
	\includegraphics[width=0.45\linewidth]{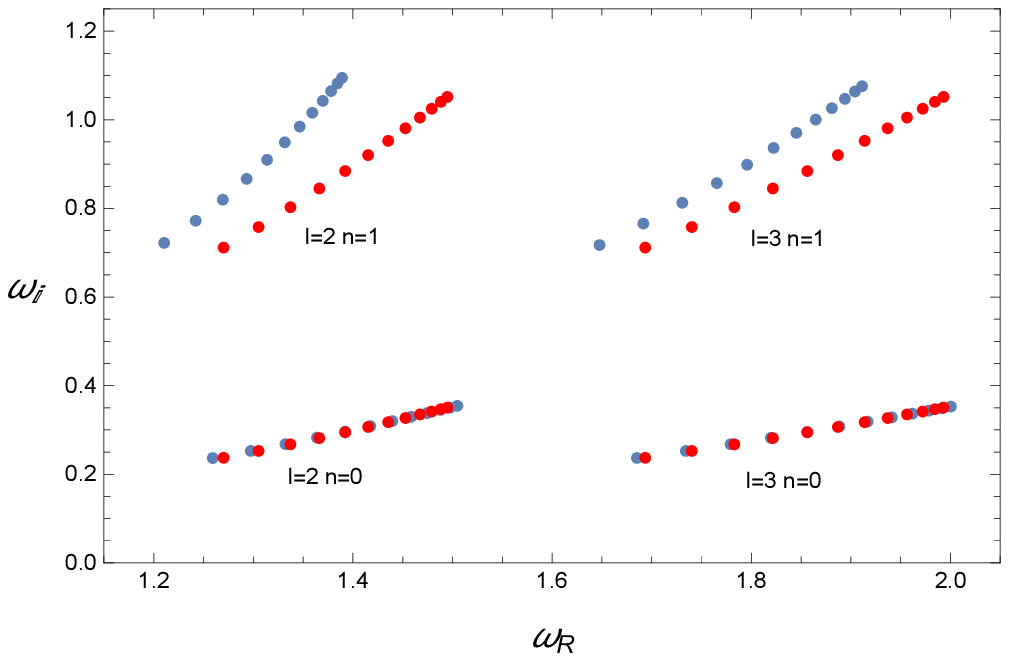} 
	\caption{The QNM frequencies of EYM black holes with $M=1$ on the $\omega_{\rm R}-\omega_{\rm I}$ plane for different values of the multiple number and the overtone number. The red dots are given by Eq.~(\ref{revreal}) and the imaginary part of Eq.~(\ref{QNM}), and the blue dots by the improved WKB method. The left diagram is devoted to the five-dimensional EYM black hole and the right to the six-dimensional one.}
	\label{fig:RI}
\end{figure}

Now we turn to investigate the modified relation for high-dimensional EYM black holes, in particular, to the case of a small multiple number. We compute the QNM frequencies by using Eq.~(\ref{revreal}) and the imaginary part of Eq.~(\ref{QNM}), and simultaneously by using the improved WKB approximation.
For the five- and six-dimensional EYM black holes, we plot the real part versus the imaginary part of QNM frequencies for different values of the multiple number and the overtone number in Fig.~\ref{fig:RI}, 
from which we find that this relation, Eq.~(\ref{revreal}), is precise enough even if the multiple number is small.\footnote{Note that the multiple number was taken to be much bigger than ours in Ref.~\cite{BCMFO} in order to keep a good agreement with the WKB method for the two models there.}
Here we present two examples, one is the case of $n=0, l=2, Q=0.5, M=1$ for the five-dimensional EYM black hole, and the other example just takes a bigger overtone number, $n=1$, but keeps the other parameters unchanged. We list the QNM frequencies obtained by the modified correspondence and the 13th order WKB method, respectively, and the relative deviation caused by the two different methods.
For the first example, the real parts are $1.366248$ and $1.363856$, respectively, and the relative deviation is $0.175\%$; 
the imaginary parts are $-0.281655$ and $-0.282719$, respectively, and the relative deviation is $0.376\%$. 
Moreover, for the second example, 
the real parts are $1.366248$ and $1.293261$, respectively, and the relative deviation is $5.644\%$;
the imaginary parts are $-0.844966$ and $-0.866700$, respectively, and the relative deviation is $2.508\%$.
We can see that the relative deviations of both real and imaginary parts will be increasing when the overtone number is bigger than zero as shown in Fig. 6.

In order to provide a sufficient support to the validity of Eq.~(\ref{revreal}), we compute more QNM frequencies and their relative deviations and list the data in six tables by taking $Q=0.1, 0.2, \dots, 0.6$, $l=0, 1,  \dots, 10, 50$, and $M=1$ for the fundamental modes ($n=0$). The reason to take such a range of $Q$ lies on the consideration that we chose the same range when we investigated the shadows of high-dimensional EYM black holes, see Fig.~\ref{fig:shadow}, that is, we maintain the coincidence of the range of $Q$. Incidentally, the data shown in Fig.~\ref{fig:RI} mainly  focus on the comparisons of QNM frequencies and their deviations to different overtone numbers.

Now we list the QNM frequencies and their relative deviations caused by the modified correspondence (Eq.~(\ref{revreal}) and the imaginary part of Eq.~(\ref{QNM})) and the improved WKB approximation for the fundamental modes in Tables~\ref{1}-\ref{6}. We make a note that the QNM frequencies for the case of $l=0$ and $D=6$ cannot be determined in terms of the WKB method, see Table~\ref{1} and Table~\ref{2}. The reason is that the calculations fail to converge to the requested accuracy for a small multiple number and a high dimension, for our case $l=0$ and $D=6$ in Table~\ref{1} and Table~\ref{2}, see Ref.~\cite{YGM,RAK2003} for the detailed discussions. We can see from the tables that the QNM frequencies computed by the former method are very close to those by the latter one. Especially for a small multiple number, see, $1\le l < 10$, the relative deviations of real parts are less than $1.510\%$. As a consequence, the data calculated by the two methods have a very good agreement even if the multiple number is small except for the cases of $l=0$, $D=6$, $Q=0.1$ and $l=0$, $D=6$, $Q=0.2$.

\begin{table}[H]
	\caption{Quasinormal mode frequencies of EYM black holes and the relative deviations caused by the modified correspondence and the improved WKB approximation, where $Q=0.1$ and $M=1$.}  
	\begin{center}  
		\scalebox{1.0}{	
		\begin{tabular}{|c|c|c|c|c|}  
			\hline  
			& \multicolumn{2}{c|}{QNM frequencies $\omega$} &\multicolumn{2}{c|}{Relative deviations} \\ \hline  
			$D=5$    & Modified correspondence &  13th WKB &${\rm Re}\,{\omega}$   & ${\rm Im}\,{\omega}$  \\ \hline  
			$l=0$ &  0.498265-0.350564$i$     & 0.530591-0.381297$i$   & 6.092\%  & 8.060\% \\  \hline  
			$l=1$ &  0.996530-0.350564$i$     & 1.011766-0.359184$i$   &1.506\%  & 2.400\% \\ \hline
			$l=2$ &  1.494795-0.350564$i$     & 1.504823-0.354418$i$   &0.666\%  & 1.087\%\\ \hline
			$l=3$ &  1.993060-0.350564$i$     & 2.000538-0.352738$i$   &0.374\%  &0.616\% \\ \hline		
			$l=4$ &  2.491325-0.350564$i$     & 2.497287-0.351955$i$   & 0.239\%  &0.395\% \\ \hline
			$l=5$ &  2.989590-0.350564$i$     & 2.994549-0.351529$i$   & 0.166\%  &0.275\% \\ \hline
			$l=10$ & 5.480916-0.350564$i$     & 5.483610-0.350850$i$   &0.049\%  & 0.082\% \\ \hline
			$l=50$ &  25.411518-0.350564$i$     & 25.412098-0.350577$i$   &0.002\%  & 0.004\% \\ \hline
			$D=6$    & Modified correspondence &  13th WKB &${\rm Re}\,{\omega}$   & ${\rm Im}\,{\omega}$  \\ \hline   
			$l=0$ &  0.844644-0.485058$i$     & ***   & ***  & *** \\  \hline  
			$l=1$ &  1.407740-0.485058$i$     & 1.426609-0.499470$i$   & 1.323\%  & 2.885\% \\ \hline
			$l=2$ &  1.970836-0.485058$i$     & 1.984219-0.492423$i$   & 0.674\%  &1.496\% \\ \hline
			$l=3$ &  2.533932-0.485058$i$     & 2.544279-0.489478$i$   & 0.407\%  &0.903\% \\ \hline		
			$l=4$ &  3.097028-0.485058$i$     & 3.105451-0.487993$i$   & 0.271\%  & 0.601\%\\ \hline
			$l=5$ &  3.660123-0.485058$i$     & 3.667223-0.487148$i$   & 0.194\%  & 0.429\% \\ \hline
			$l=10$ & 6.475603-0.485058$i$    & 6.479581-0.485717$i$    & 0.061\%  & 0.136\%\\ \hline
			$l=50$ &  28.999440-0.485058$i$   & 29.000324-0.485090$i$   &0.003\%  &0.007\%\\ \hline 
		\end{tabular}  
	}
	\end{center}\label{1}  
\end{table} 

\begin{table}[H]
	\caption{Quasinormal mode frequencies of EYM black holes and the relative deviations caused by the modified correspondence and the improved WKB approximation, where $Q=0.2$ and $M=1$.}  
	\begin{center}  
		\scalebox{1.0}{	
		\begin{tabular}{|c|c|c|c|c|}  
			\hline  
			& \multicolumn{2}{c|}{QNM frequencies $\omega$} &\multicolumn{2}{c|}{Relative deviations} \\ \hline  
			$D=5$    & Modified correspondence &  13th WKB &${\rm Re}\,{\omega}$   & ${\rm Im}\,{\omega}$  \\ \hline   
			$l=0$ &  0.493035-0.341642$i$     &0.520853-0.370485$i$   & 5.341\%  & 7.785\% \\  \hline  
			$l=1$ &  0.986071-0.341642$i$     & 0.998915-0.349397$i$   &1.286\%  & 2.220\% \\ \hline
			$l=2$ &  1.479106-0.341642$i$     & 1.487527-0.345118$i$   &0.566\%  & 1.007\% \\ \hline
			$l=3$ &  1.972142-0.341642$i$     & 1.978402-0.343594$i$   &0.316\%  & 0.568\% \\ \hline		
			$l=4$ &  2.465177-0.341642$i$     & 2.470160-0.342887$i$   &0.202\%  & 0.363\% \\ \hline
			$l=5$ &  2.958213-0.341642$i$     & 2.962352-0.342504$i$   &0.140\%  & 0.252\% \\ \hline
			$l=10$ & 5.423390-0.341642$i$     & 5.425635-0.341897$i$   &0.041\%  & 0.075\% \\ \hline
			$l=50$ & 25.144808-0.341642$i$   & 25.145291-0.341654$i$   &0.002\% &  0.004\% \\ \hline
			$D=6$    & Modified correspondence &  13th WKB &${\rm Re}\,{\omega}$   & ${\rm Im}\,{\omega}$  \\ \hline  
			$l=0$ &  0.812255-0.459507$i$     & ***   & ***  & *** \\ \hline  
			$l=1$ &  1.353758-0.459507$i$     & 1.370386-0.472319$i$   & 1.213\%  & 2.713\% \\ \hline
			$l=2$ &  1.895262-0.459507$i$     & 1.907037-0.466049$i$   & 0.617\%  & 1.404\% \\ \hline
			$l=3$ &  2.436765-0.459507$i$     & 2.445863-0.463432$i$   & 0.372\%  & 0.847\%\\ \hline		
			$l=4$ &  2.978269-0.459507$i$     & 2.985673-0.462115$i$   & 0.248\%  & 0.564\%\\ \hline
			$l=5$ &  3.519772-0.459507$i$     & 3.526012-0.461364$i$   & 0.177\%  & 0.403\% \\ \hline
			$l=10$ & 6.227289-0.459507$i$    & 6.230785-0.460092$i$   & 0.056\%  &  0.127\%	\\ \hline
			$l=50$ &  27.887424-0.459507$i$   & 27.888201-0.459536$i$   &0.003\%  & 0.006\%\\ \hline 
		\end{tabular}  
	}
	\end{center}\label{2}  
\end{table}

\begin{table}[H]
	\caption{Quasinormal mode frequencies of EYM black holes and the relative deviations caused by the modified correspondence and the improved WKB approximation, where $Q=0.3$ and $M=1$.}  
	\begin{center}  
		\scalebox{1.0}{	
		\begin{tabular}{|c|c|c|c|c|}  
			\hline  
			& \multicolumn{2}{c|}{QNM frequencies $\omega$} &\multicolumn{2}{c|}{Relative deviations} \\ \hline  
			$D=5$    & Modified correspondence &  13th WKB &${\rm Re}\,{\omega}$   & ${\rm Im}\,{\omega}$  \\ \hline  
			$l=0$ &  0.484237-0.326938$i$     & 0.504477-0.351362$i$   & 4.012\%  &6.951\% \\  \hline  
			$l=1$ &  0.968474-0.326938$i$     & 0.977465-0.333452$i$   & 0.920\% & 1.954\% \\ \hline
			$l=2$ &  1.452711-0.326938$i$     & 1.458492-0.329802$i$   & 0.396\%  &0.868\% \\ \hline
			$l=3$ &  1.936949-0.326938$i$     & 1.941209-0.328530$i$   &0.219\%  &0.485\% \\ \hline		
			$l=4$ &  2.421186-0.326938$i$     & 2.424561-0.327949$i$   & 0.139\% &0.308\% \\ \hline
			$l=5$ &  2.905423-0.326938$i$     & 2.908218-0.327636$i$   & 0.096\%  &0.213\% \\ \hline
			$l=10$ & 5.326608-0.326938$i$     & 5.328116-0.327144$i$   &0.028\%  &0.063\% \\ \hline
			$l=50$ & 24.696094-0.326938$i$   & 24.696417-0.326948$i$   &0.001\% &0.003\% \\ \hline
			$D=6$    & Modified correspondence &  13th WKB &${\rm Re}\,{\omega}$   & ${\rm Im}\,{\omega}$  \\ \hline    
			$l=0$ &  0.764095-0.422875$i$     & 0.787939-0.451020$i$ & 3.026\%   & 6.240\% \\ \hline  
			$l=1$ &  1.273491-0.422875$i$     & 1.287106-0.433584$i$   &1.058\%  &2.470\% \\ \hline
			$l=2$ &  1.782888-0.422875$i$     & 1.792500-0.428329$i$   & 0.536\%  & 1.273\% \\ \hline
			$l=3$ &  2.292284-0.422875$i$     & 2.299709-0.426152$i$   &0.323\%  & 0.769\%\\ \hline		
			$l=4$ &  2.801681-0.422875$i$     & 2.807721-0.425052$i$   &0.215\%  & 0.512\%\\ \hline
			$l=5$ &  3.311077-0.422875$i$     & 3.316166-0.424426$i$   & 0.153\%  & 0.365\% \\ \hline
			$l=10$ & 5.858059-0.422875$i$     & 5.860910-0.423365$i$   & 0.049\%  & 0.116\%	\\ \hline
			$l=50$ &  26.233918-0.422875$i$   & 26.234551-0.422899$i$   &0.002\%  & 0.006\%\\ \hline 
		\end{tabular}  
	}
	\end{center}\label{3}  
\end{table}

\begin{table} [H]
	\caption{Quasinormal mode frequencies of EYM black holes and the relative deviations caused by the modified correspondence and the improved WKB approximation, where $Q=0.4$ and $M=1$.}  
	\begin{center}  
		\scalebox{1.0}{	
		\begin{tabular}{|c|c|c|c|c|}  
			\hline  
			& \multicolumn{2}{c|}{QNM frequencies $\omega$} &\multicolumn{2}{c|}{Relative deviations} \\ \hline  
			$D=5$    & Modified correspondence &  13th WKB &${\rm Re}\,{\omega}$   & ${\rm Im}\,{\omega}$  \\ \hline  
			$l=0$ &  0.471750-0.306757$i$     & 0.482955-0.330156$i$   & 2.320\%& 7.087\% \\  \hline  
			$l=1$ &  0.943499-0.306757$i$     & 0.947146-0.311584$i$   & 0.390\%& 1.549\%  \\ \hline
			$l=2$ &  1.415249-0.306757$i$     & 1.417409-0.308803$i$   & 0.152\%& 0.663\%  \\ \hline
			$l=3$ &  1.886998-0.306757$i$     & 1.888518-0.307872$i$   & 0.080\%& 0.362\% \\ \hline
			$l=4$ &  2.358748-0.306757$i$     & 2.359919-0.307457$i$   & 0.050\% & 0.228\% \\ \hline
			$l=5$ &  2.830497-0.306757$i$     & 2.831450-0.307237$i$   & 0.034\% & 0.156\% \\ \hline
			$l=10$ &  5.189245-0.306757$i$    & 5.189743-0.306896$i$   & 0.010\%& 0.045\% \\ \hline
			$l=50$ &  24.059228-0.306757$i$   & 24.059334-0.306763$i$  & 0.000\% & 0.002\% \\ \hline
			$D=6$    & Modified correspondence &  13th WKB &${\rm Re}\,{\omega}$   & ${\rm Im}\,{\omega}$  \\ \hline  
			$l=0$ &  0.706829-0.381442$i$     & 0.725290-0.404194$i$   & 2.545\%& 5.629\% \\  \hline  
			$l=1$ &  1.178049-0.381442$i$     & 1.188558-0.390026$i$   & 0.884\%& 2.201\%  \\ \hline
			$l=2$ &  1.649268-0.381442$i$     & 1.656669-0.385809$i$   & 0.447\%& 1.132\%  \\ \hline
			$l=3$ &  2.120488-0.381442$i$     & 2.126203-0.384051$i$   & 0.269\%& 0.679\% \\ \hline
			$l=4$ &  2.591707-0.381442$i$     & 2.596348-0.383172$i$   & 0.179\% & 0.451\% \\ \hline
			$l=5$ &  3.062927-0.381442$i$     & 3.066837-0.382670$i$   & 0.127\% &0.321\% \\ \hline
			$l=10$ & 5.419024-0.381442$i$    & 5.421214-0.381817$i$    & 0.040\%& 0.098\% \\ \hline
			$l=50$ &  24.267804-0.381442$i$   & 24.268290-0.381442$i$  & 0.002\% & 0.000\% \\ \hline
		\end{tabular} 
	} 
	\end{center}  \label{4}
\end{table} 

\begin{table}[H]
	\caption{Quasinormal mode frequencies of EYM black holes and the relative deviations caused by the modified correspondence and the improved WKB approximation, where $Q=0.5$ and $M=1$.}
	\begin{center}  
		\scalebox{1.0}{	
		\begin{tabular}{|c|c|c|c|c|}  
			\hline  
			& \multicolumn{2}{c|}{QNM frequencies $\omega$} &\multicolumn{2}{c|}{Relative deviations} \\ \hline  
			$D=5$    & Modified correspondence &  13th WKB &${\rm Re}\,{\omega}$   & ${\rm Im}\,{\omega}$  \\ \hline  
			$l=0$ &  0.455416-0.281655$i$     & 0.452094-0.295107$i$   & 0.735\%  & 4.558\% \\  \hline  
			$l=1$ &  0.910832-0.281655$i$     & 0.907866-0.284435$i$   & 0.327\%  & 0.944\% \\ \hline
			$l=2$ &  1.366248-0.281655$i$     & 1.363856-0.282719$i$   & 0.175\%  & 0.376\% \\ \hline
			$l=3$ &  1.821665-0.281655$i$     & 1.819739-0.282202$i$   & 0.106\%  & 0.194\% \\ \hline		
			$l=4$ &  2.277081-0.281655$i$     & 2.275482-0.281986$i$   & 0.070\%  & 0.117\% \\ \hline
			$l=5$ &  2.732497-0.281655$i$     & 2.731136-0.281878$i$   & 0.050\%  & 0.079\% \\ \hline
			$l=10$ &  5.009578-0.281655$i$     & 5.008809-0.281718$i$   & 0.015\%  & 0.022\% \\ \hline
			$l=50$ &  23.226224-0.281655$i$     & 23.226056-0.281658$i$   & 0.001\%  & 0.001\% \\ \hline
			$D=6$    & Modified correspondence &  13th WKB&${\rm Re}\,{\omega}$   & ${\rm Im}\,{\omega}$  \\ \hline   
			$l=0$ &  0.646739-0.340336$i$     & 0.660448-0.358480$i$   & 2.076\%  & 5.061\% \\  \hline  
			$l=1$ &  1.077899-0.340336$i$     & 1.085679-0.347140$i$   & 0.717\%  & 1.960\% \\ \hline
			$l=2$ &  1.509058-0.340336$i$     & 1.514521-0.343802$i$   & 0.361\%  & 1.008\% \\ \hline
			$l=3$ &  1.940218-0.340336$i$     & 1.944424-0.342417$i$   & 0.216\%  & 0.608\% \\ \hline		
			$l=4$ &  2.371377-0.340336$i$     & 2.374795-0.341722$i$   & 0.144\%  & 0.406\% \\ \hline
			$l=5$ &  2.802537-0.340336$i$     & 2.805415-0.341324$i$   & 0.103\%  & 0.289\% \\ \hline
			$l=10$ &  4.958334-0.340336$i$    & 4.959945-0.340649$i$   & 0.032\%  & 0.092\% \\ \hline
			$l=50$ &  22.204713-0.340336$i$   & 22.205071-0.340351$i$   & 0.002\%  & 0.004\% \\ \hline 
		\end{tabular}  
	}
	\end{center}  \label{5}
\end{table} 

\begin{table}[H]
	\caption{Quasinormal mode frequencies of EYM black holes and the relative deviations caused by the modified correspondence and the improved WKB approximation, where $Q=0.6$ and $M=1$.}  
	\begin{center}  
		\scalebox{1.0}{	
		\begin{tabular}{|c|c|c|c|c|}  
			\hline  
			& \multicolumn{2}{c|}{QNM frequencies $\omega$} &\multicolumn{2}{c|}{Relative deviations} \\ \hline  
			$D=5$    & Modified correspondence &  13th WKB &${\rm Re}\,{\omega}$   & ${\rm Im}\,{\omega}$  \\ \hline  
			$l=0$ &  0.435077-0.252648$i$     & 0.416501-0.258893$i$   & 4.460\%  & 2.412\% \\  \hline  
			$l=1$ &  0.870155-0.252648$i$     & 0.859046-0.253097$i$   & 1.293\%  & 0.177\% \\ \hline
			$l=2$ &  1.305232-0.252648$i$     & 1.297374-0.252624$i$   & 0.606\%  & 0.010\% \\ \hline
			$l=3$ &  1.740310-0.252648$i$     & 1.734254-0.252572$i$   &0.349\%  &0.030\% \\ \hline		
			$l=4$ &  2.175387-0.252648$i$     & 2.170474-0.252579$i$   & 0.226\%  &0.024\% \\ \hline
			$l=5$ &  2.610465-0.252648$i$     & 2.606337-0.252591$i$   & 0.158\%  & 0.023\% \\ \hline
			$l=10$ &  4.785852-0.252648$i$     & 4.783569-0.252627$i$   & 0.048\% & 0.007\% \\ \hline
			$l=50$ &  22.188951-0.252648$i$     & 22.188455-0.252647$i$   &0.002\%  & 0.000\% \\ \hline
			$D=6$    & Modified correspondence &  13th WKB&${\rm Re}\,{\omega}$   & ${\rm Im}\,{\omega}$  \\ \hline   
			$l=0$ &  0.588483-0.302769$i$     & 0.598358-0.317238$i$   & 1.650\%  & 4.561\% \\  \hline  
			$l=1$ &  0.980805-0.302769$i$     & 0.986426-0.308174$i$   & 0.570\%  & 1.754\% \\ \hline
			$l=2$ &  1.373128-0.302769$i$     & 1.377058-0.305522$i$   & 0.285\%  &0.901\% \\ \hline
			$l=3$ &  1.765450-0.302769$i$     & 1.768471-0.304425$i$   & 0.171\%  &0.544\% \\ \hline		
			$l=4$ &  2.157772-0.302769$i$     & 2.160225-0.303873$i$   & 0.114\%  & 0.363\% \\ \hline
			$l=5$ &  2.550094-0.302769$i$     & 2.552159-0.303557$i$   &0.081\%  & 0.260\%\\ \hline
			$l=10$ & 4.511705-0.302769$i$    & 4.512860-0.303019$i$   &0.026\%  &0.083\% \\ \hline
			$l=50$ &  20.204590-0.302769$i$   & 20.204847-0.302782$i$   & 0.001\% & 0.004\%\\ \hline 
		\end{tabular}
	}  
	\end{center}\label{6}  
\end{table}

\section{Conclusions}\label{sec:conclusions}
We have studied the null geodesics and the correspondence between QNMs in the eikonal limit and shadow radii for high-dimensional EYM black holes. For the five-dimensional case, two branches of the radius of a photon sphere exist, but only the one outside the event horizon satisfies the conditions of circular null geodesics. In general, the radii of photon spheres are mass independent when the gauge charge is large, which is same as the behavior of the horizon radii. In addition,
the reflecting point of shadow radii and the spiral-like shape on the $\Omega-\lambda$ plane do not appear in high-dimensional EYM spacetimes.
In particular, for the modified correspondence between the real part of QNMs in the eikonal limit and the shadow radius, Eq.~(\ref{revreal}), we verify its validity and find that it has a better agreement to the WKB method for a larger multiple number but a smaller overtone number in high-dimensional EYM black holes.

Finally, we emphasize that the modified correspondence, Eq.~(\ref{revreal}), has a good enough agreement to the WKB method for the fundamental modes with a vanishing overtone number in high-dimensional Einstein-Yang-Mills spacetimes even if the multiple number is small. The significance of such a consequence is that we establish a deep connection between gravitational waves and shadows of black holes. As is known, the fundamental modes always dominate the waveform of gravitational waves because of their least damped feature in a detected ringdown signal.
Although we only discuss the QNMs of a scalar field perturbation in this paper, the QNMs embody the intrinsic property of black holes and they are determined by the parameters of black holes, such as mass, charge, etc., and independent of the type of perturbation fields. 
Our finding might be understood as an exploration of the underlying feature of the correspondence between gravitational waves and shadows of black holes.

\section*{Acknowledgments}

The authors would like to thank the anonymous referee for the helpful comments that improve this work greatly. This work was supported in part by the National Natural Science Foundation of China under grant No. 11675081.

\end{document}